\documentstyle[epsf,preprint,twoside,adbis97%
,amscd%
,times]{acmconf} 

\author{L.Yu.Ismailova, K.E. Zinchenko \vspace{1.52mm} \\
{Vorotnikovskiy per, 7, bld. 4} \\
{JurInfoR-MSU Institute for Contemporary Education} \\
{Moscow, 103006, Russia}\\
{\tt \{larisa,kz\}@jurinfor.ru} \\
{\tt WWW: http://www.jurinfor.ru}}
\addtocounter{footnote}{1}
\title{An object evaluator to generate flexible applications \\%
(Extended abstract)
}

\begin{document}

\bibliographystyle{alpha}

\maketitle


\begin{abstract}
{\small This paper contains a brief discussion of an object
evaluator which is based on principles of evaluations in a
category. The main tool system referred as the Application
Development Environment (ADE) is used to build database
applications involving the graphical user interface (GUI). The
separation of a database access and the user interface is reached
by distinguishing the potential and actual objects. The variety of
applications may be generated that communicate with different and
distinct desktop databases. The commutative diagrams' technique
allows to involve retrieval and call of the delayed procedures. }
\end{abstract}


\section*{Introduction}

Recent research activity generated the valuable advance in
understanding the nature of an {\em object} withing the information
system. Filling in the gap between the manifold
of known data models, implemented information systems and
theoretical approaches caused
the experimental efforts in
development the object-oriented tools.

Here is breafly discussed the Application Development Environment (ADE)
that is used to build database applications involving the graphical user
interface (GUI).
ADE computing separates the database access and the user
interface. The variety of applications may be generated that communicate
with different and distinct desktop databases. The advanced techniques
allows to involve remote or stored procedures retrieval and call.

According to an object-oriented traditions
\cite{Report:96:Frederiks:OO-Arch},
ADE include some basic features
of inheritance, encapsulation, and polymorphism. They are used to derive an
actual object to cover the needed information resources.

The {\em potential object} (PO) is composed with the {\em menu} (M),
{\em data access} (DA),
and {\em modular counterparts} (MC).
The {\em Ancestor Potential Objects} (APO) contain
the menus, events, event evolver, attributes and functions (that are
encapsulated). The {\em Descendant Potential Objects} (DPO) are inherited from
APO.

The aim of the current contribution is to give a brief profile
of the ADE project without any detailed mathematical
or implementational consideration. Nevertheless, some mathematical
background corresponds to the references
\cite{Book:95:Abiteboul:DBFundamentals},
\cite{Article:96:Ruiz:OOFormMeth},
\cite{Article:96:Fitzgerald:FormMethodologies}.
Other less traditional
for the database area ideas are due to
\cite{Article:93:Dampney:AMAST93}
to conform the {\em object computation} strategies
using the {\em commutative diagrams}. The main ADE
building blocks have the relative uniformity to resolve
the modular linkages. ADE enables the host computational
environment to extend the properties of the distinct MC.

\section{Indexing the objects by events}\label{Section:1}

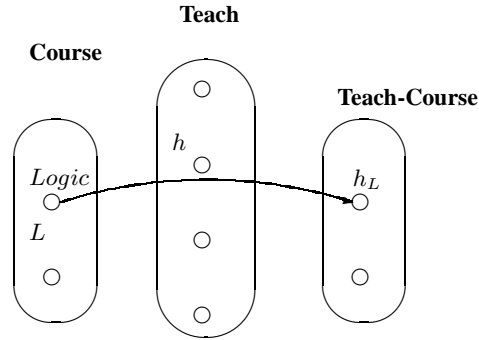
\begin{figure}
\unitlength 1.00mm \linethickness{0.4pt}
\begin{picture}(53.00,44.00)
\put(23.00,44.00){\makebox(0,0)[lc]{\bf Teach}}
\put(26.00,34.00){\circle{2.00}} \put(26.00,24.00){\circle{2.00}}
\put(26.00,14.00){\circle{2.00}}
\put(22.00,27.00){\makebox(0,0)[lc]{$h$}}
\put(3.00,39.00){\makebox(0,0)[lc]{\bf Course}}
\put(6.00,19.00){\circle{2.00}} \put(6.00,9.00){\circle{2.00}}
\put(3.00,22.00){\makebox(0,0)[lc]{$Logic$}}
\put(44.00,33.00){\makebox(0,0)[lc]{\bf Teach-Course}}
\put(47.00,19.00){\circle{2.00}} \put(47.00,9.00){\circle{2.00}}
\put(46.00,22.00){\makebox(0,0)[lc]{$h_L$}}
\put(3.00,15.00){\makebox(0,0)[lc]{$L$}}
\put(6.50,16.50){\oval(11.00,27.00)[]}
\put(26.00,4.00){\circle{2.00}}
\put(47.50,16.50){\oval(11.00,27.00)[]}
\put(26.50,19.50){\oval(13.00,37.00)[]}
\put(46.00,19.00){\vector(4,-1){0.2}}
\multiput(7.00,19.00)(0.38,0.11){6}{\line(1,0){0.38}}
\multiput(9.28,19.69)(0.46,0.12){5}{\line(1,0){0.46}}
\multiput(11.58,20.28)(0.46,0.10){5}{\line(1,0){0.46}}
\multiput(13.89,20.79)(0.58,0.10){4}{\line(1,0){0.58}}
\multiput(16.21,21.21)(0.78,0.11){3}{\line(1,0){0.78}}
\multiput(18.54,21.54)(0.78,0.08){3}{\line(1,0){0.78}}
\multiput(20.88,21.78)(1.18,0.08){2}{\line(1,0){1.18}}
\put(23.24,21.94){\line(1,0){2.37}}
\put(25.61,22.00){\line(1,0){2.38}}
\put(27.99,21.97){\line(1,0){2.39}}
\multiput(30.39,21.86)(1.20,-0.10){2}{\line(1,0){1.20}}
\multiput(32.79,21.65)(0.81,-0.10){3}{\line(1,0){0.81}}
\multiput(35.21,21.36)(0.61,-0.10){4}{\line(1,0){0.61}}
\multiput(37.64,20.97)(0.61,-0.12){4}{\line(1,0){0.61}}
\multiput(40.09,20.50)(0.49,-0.11){5}{\line(1,0){0.49}}
\multiput(42.54,19.94)(0.43,-0.12){8}{\line(1,0){0.43}}
\end{picture}
\caption{\sf {\bf E}vent {\bf D}riven {\bf O}bjects. {\small
(Here: Let event be an assigning {\em potential} teachers to the
courses. E.g., $Logic$ is under assigning so that potential
teachers become the {\em actual} teachers (of $Logic$). The domain
{\bf Course} is used as a category of indices, {\bf Teach} is a
domain of possible teachers (not assigned to courses), and {\bf
Teach-Course} is a domain of the subsets of {\bf Teach}, each
subset being corresponded to the specifies index. More generally,
possible object $h$ is a map from the event (index) $L$ into the
actual object $h_L$.) } }\label{pic:1}
\end{figure}

The main feature to evaluate the objects within modular
structures is based on the triggering events.
An event triggers the assignment of potential objects
to the domains of the actual objects. All of this
is done as a singular Modular Counterpart (MC).
Each MC is attached to object which is thus determined.
The MC is a holder of all the controls to communicate with the user. The
event is assigned by the user call (for instance, clicking) or selection. Thus,
when the activity is initiated, the following main events may be triggered:
respond to a request from the user application, database retrieval or
updating. The possible order of the events is prescribed by {\em evolver}
and is detrmined by the {\em scripts}. A fragment of the event
driven procedure is shown in Figure~\ref{pic:1}.

Note that a set of the possible objects
$\{h|h:Course \to Teasch-Logic\}=H_{Teach-Logic}(Course)$
in this figure
represents an idea of functor-as-object for $Course$ is a category
of events, $Teach-Logic$ is a (sub)category
of the actual objects -- {\em type}.
More rigorously, $h_L$ belongs to  a type {\em derived}
as followed: possible object $h$ from the domain {\bf Teach}
(i.e. it has this {\em sort}) is applied to the index $L$
(i.e. is observed from $L$), so that the event $L$
triggered the object $h_L$ to be born.

\subsection{Triggering events: menu}

Menu gives more flexibility to the attribute selection. Ususally the lists of
possible attributes are supported to give the developer or user more freedom.
Menus are established to be encapsulated in APO and are inherited in DPO.

\begin{figure}
\unitlength 1.00mm \linethickness{0.4pt}
\begin{picture}(80.00,140.00)
\put(10.00,65.00){\framebox(20.00,7.00)[cc]{$Informatics$}}
\put(10.00,72.00){\framebox(20.00,7.00)[cc]{$Informatics$}}
\put(10.00,79.00){\framebox(20.00,7.00)[cc]{$Logic$}}
\put(10.00,86.00){\framebox(20.00,7.00)[cc]{$Logic$}}
\put(30.00,65.00){\framebox(20.00,7.00)[cc]{$Jackson$}}
\put(30.00,72.00){\framebox(20.00,7.00)[cc]{$Doe$}}
\put(30.00,79.00){\framebox(20.00,7.00)[cc]{$Smith$}}
\put(30.00,86.00){\framebox(20.00,7.00)[cc]{$Johnes$}}
\put(50.00,65.00){\framebox(10.00,7.00)[cc]{$30$}}
\put(50.00,72.00){\framebox(10.00,7.00)[cc]{$30$}}
\put(50.00,79.00){\framebox(10.00,7.00)[cc]{$20$}}
\put(50.00,86.00){\framebox(10.00,7.00)[cc]{$20$}}
\put(20.00,95.00){\makebox(0,0)[cc]{$Course$}}
\put(40.00,95.00){\makebox(0,0)[cc]{$Name$}}
\put(55.00,95.00){\makebox(0,0)[cc]{$Hours$}}
\put(30.00,105.00){\framebox(20.00,7.00)[cc]{$Jackson$}}
\put(30.00,112.00){\framebox(20.00,7.00)[cc]{$Doe$}}
\put(30.00,119.00){\framebox(20.00,7.00)[cc]{$Smith$}}
\put(30.00,126.00){\framebox(20.00,7.00)[cc]{$Johnes$}}
\put(40.00,135.00){\makebox(0,0)[cc]{$Name$}}
\put(5.00,119.00){\framebox(20.00,7.00)[cc]{$Informatics$}}
\put(5.00,126.00){\framebox(20.00,7.00)[cc]{$Logic$}}
\put(15.00,135.00){\makebox(0,0)[cc]{$Course$}}
\put(55.00,119.00){\framebox(10.00,7.00)[cc]{$30$}}
\put(55.00,126.00){\framebox(10.00,7.00)[cc]{$20$}}
\put(60.00,135.00){\makebox(0,0)[cc]{$Hours$}}
\put(60.00,34.00){\framebox(20.00,7.00)[cc]{$Jackson$}}
\put(60.00,41.00){\framebox(20.00,7.00)[cc]{$Doe$}}
\put(55.00,56.00){\makebox(0,0)[lc]{\bf Teach-Informatics:}}
\put(60.00,50.00){\makebox(0,0)[lc]{$Tch_{Inf}$}}
\put(5.00,140.00){\makebox(0,0)[lc]{\bf Domains:}}
\put(5.00,100.00){\makebox(0,0)[lc]{\bf Relationship-1:}}
\put(32.00,15.00){\framebox(20.00,7.00)[cc]{$Jackson$}}
\put(32.00,22.00){\framebox(20.00,7.00)[cc]{$Doe$}}
\put(32.00,29.00){\framebox(20.00,7.00)[cc]{$Smith$}}
\put(32.00,36.00){\framebox(20.00,7.00)[cc]{$Johnes$}}
\put(27.00,51.00){\makebox(0,0)[lc]{\bf Teach:}}
\put(32.00,45.00){\makebox(0,0)[lc]{$Tch$}}
\put(60.00,2.00){\framebox(20.00,7.00)[cc]{$Smith$}}
\put(60.00,9.00){\framebox(20.00,7.00)[cc]{$Johnes$}}
\put(55.00,24.00){\makebox(0,0)[lc]{\bf Teach-Logic:}}
\put(60.00,18.00){\makebox(0,0)[lc]{$Tch_L$}}
\put(5.00,43.00){\framebox(20.00,7.00)[cc]{$Informatics$}}
\put(5.00,50.00){\framebox(20.00,7.00)[cc]{$Logic$}}
\put(5.00,60.00){\makebox(0,0)[lc]{\bf Course:}}
\put(12.00,26.00){\vector(3,-2){32.00}}
\put(18.00,11.00){\makebox(0,0)[cc]{Indexing}}
\end{picture}
\caption{ {\sf Indexing the {\em actual} objects.} {\small (Here:
{\bf Teach} is the {\em intensional} object (denoted by $Tch$) and
its {\em extention} correspondes to the domain of {\em potential}
objects -- the professors are the potential teachers. The {\em
actual} objects are established after {\em indexing}. Let the
domain of indices be the {\bf Course}. The domains {\bf
Teach-Logic} (denoted by $Tch_{L}$ and {\bf Teach-Informatic}
(denoted by $Tch_{Inf}$) are established as the {\em actual}
objects -- the professors are the actual teachers and assigned to
the courses. On an other hand {\bf Relationship-1} is a plain
relation which gives an assignment of professors to the courses
and courses to the hours, and all of this is in accordance with
the normalized relational model. Note that this relationship gives
the local univers in spite of the potential nature of its tuples
and their elements. From this point of view the indexing captures
more meaning.)} }\label{pic:3}
\end{figure}
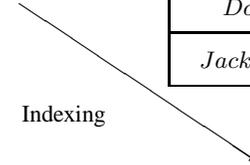

An exmaple could assist the ideas beyond the menu driven objects.
In Figure~\ref{pic:3} the events of assigning teachers to courses
reflect a pure relational solution (by plain relations)
and the enhanced solution. The enhancement is done by
observing the triggering event as a stimulus to generate
the actual object.
The most prominent feature is that
the {\em actual} objects are established after
{\em indexing}. It means that before indexing
there are no visible actual objects, and all the
objects are potential.
Let the domain of indices be the {\bf Course}.
The domains {\bf Teach-Logic} (denoted by $Tch_{L}$
and {\bf Teach-Informatic} (denoted by $Tch_{Inf}$)
are established as the {\em actual} objects -- the professors
are the actual teachers and assigned to the courses.
Another way to construe and maintain the relationship
between $Course$, $Name$ and $Hours$ is given
by their direct correspondence.
In the Figure~\ref{pic:3} the table of
instances {\bf Relationship-1} is a plain
relation which gives an assignment of professors to the courses
and courses to the hours, and all of this is in accordance
with the normalized relational model. Note that this relationship
gives the local univers in spite of the potential nature
of its tuples and their elements. From this point of view the
indexing captures more meaning.

Thus, clicking the menu {\bf Course} user triggers an assigning
process which result in the {\bf Teach-Course} domain.

\subsection{Deriving types in a particular application}

\begin{figure*}
\unitlength 1.00mm \linethickness{0.4pt}
\begin{picture}(131.00,38.00)
\put(24.00,33.00){\framebox(15.00,5.00)[cc]{$Jones$}}
\put(24.00,28.00){\framebox(15.00,5.00)[cc]{$Smith$}}
\put(13.00,35.00){\makebox(0,0)[lc]{$Logic\ ($}}
\put(41.00,35.00){\makebox(0,0)[lc]{$/x)$}}
\put(68.00,35.00){\vector(-1,0){19.00}}
\put(83.00,33.00){\framebox(15.00,5.00)[cc]{$Jones$}}
\put(83.00,28.00){\framebox(15.00,5.00)[cc]{$Smith$}}
\put(70.00,35.00){\makebox(0,0)[lc]{\normalsize $[Logic,\ $}}
\put(101.00,35.00){\makebox(0,0)[cc]{\normalsize$]$}}
\put(104.00,35.00){\vector(1,0){24.00}}
\put(54.00,17.00){\makebox(0,0)[lc]{$\hat{}\ Tch \times
id_{Name}$}} \put(131.00,35.00){\makebox(0,0)[lc]{$true$}}
\put(13.00,17.00){\makebox(0,0)[lc]{$\hat{}\ Tch$}}
\put(22.00,4.00){\makebox(0,0)[lc]{\normalsize $Tch_L$}}
\put(83.00,7.00){\framebox(15.00,5.00)[cc]{$Jones$}}
\put(83.00,2.00){\framebox(15.00,5.00)[cc]{$Smith$}}
\put(70.00,4.00){\makebox(0,0)[lc]{$[Tch_L,\ $}}
\put(101.00,4.00){\makebox(0,0)[cc]{$]$}}
\put(53.00,6.00){\makebox(0,0)[cc]{$Fst$}}
\put(59.00,37.00){\makebox(0,0)[cc]{$Subst_x$}}
\put(111.00,17.00){\makebox(0,0)[cc]{$\varepsilon$}}
\put(115.00,37.00){\makebox(0,0)[cc]{$Tch-Filter$}}
\put(104.00,5.00){\vector(1,1){25.00}}
\put(76.00,30.00){\vector(0,-1){22.00}}
\put(25.00,26.00){\vector(0,-1){18.00}}
\put(66.00,4.00){\vector(-1,0){36.00}}
\put(24.00,26.00){\line(1,0){2.00}}
\put(68.00,34.00){\line(0,1){2.00}}
\put(66.00,5.00){\line(0,-1){2.00}}
\put(75.00,30.00){\line(1,0){2.00}}
\put(104.00,34.00){\line(0,1){2.00}}
\put(104.00,6.00){\line(0,-1){2.00}}
\end{picture}
\caption{ {\sf Genration of the {\em actual} objects by indexing.}
{\small (Here: This commutative diagram has an entry point $Logic$
and the exit point $Tch_L$ and is supplementary to the domains in
Figure~\ref{pic:3}. The element $Logic$ is from the domain of
indices {\bf Course}, and the domain of potential objects {\bf
Teach} is given relatively {\bf Course}. This relativisation
generates the domain of actual objects $Tch_L$, or {\bf
Teach-Logic}. This separation is due to the logical filtering
$Tch-Filter$ of the input objects. The filtering is composed from
the substitutional part $Subst_x$ and map $\hat{}\ Tch$. The
substitution $Subst_x$ modifies the computational environment so
that, e.g., $Logic(Jones/x)$ means that the occurence of variable
$x$ is replaced by the element $Jones$ being tested. Whenever the
positive result is generated the element is pushed into the domain
$Tch_L$ as its new inhabitant. The notation $\hat{}\ Tch$ is used
to mark the shifting of index and $Fst$ is the first projection of
the ordered pair being tested. The map $\varepsilon$ applies the
function object to an argument object and results in the function
value.)} }\label{pic:4}
\end{figure*}
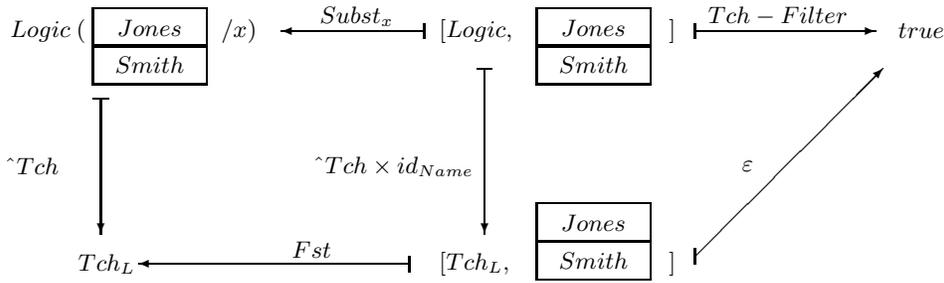

The particular application is derived from Potential Object Library (POL)
giving rise to Actual Object Libraries (AOL). A computational idea
in use is given in Figure~\ref{pic:4}.
The correspondence between potential and actual objects
is established in a category style evaluation. Note,
that the known effects to generate the value of
a logical expression tend to construe the {\em filters}.
Their mathematical properties are left out of the scope
of this paper. A feature of the filter is to compose
its substitutional part with the evaluation part.
The substitution effects an environment so that
an {\em old} environment is mapped into the {\em new}
environment. The modified environment contains the
{\em encapsulated} instances under a further evaluation.

\section{Computational and relative backgrounds}\label{Section:2}

ADE concepts are based on a variant of computation theory for information
systems. The main idea is to observe the function and an
argumrnt as the distinct and separate entities.
The function value is generated whenever the function object
is {\em applied to} an argument object.
During the evaluation process the old environment
is mapped into the new environment. This transition
captures the necessary inclusions, e.g., encpsulated
partiotions of the computation.
The switching of an environment gives the effect of transition
from a `previous stage of knowledge' to the
`current stage of knowledge' resulting in the evolution
of the concepts. Their computational solutions are fitted to
the known object-oriented paradigms, also data base and knowledge base
engineering are included.

Some vital concepts have clear mathematical
representions: {\em data object} that represents the computer stored data;
{\em metadata object} that represents the conceptual information;
{\em assignment} that captures more dynamics and intrinsic states;
{\em expansible database} that represent the
individualized self contained object couples.

The main feature of the general
approach is in embedding the typed entities into
the initially typeless system that is based on the a theory of the
potential, or possible objects.
The last one is fitted to capture the dynamics of different objects and
switching the states in the information systems.
The prescribed modes in the
information systems as a rule are corresponded to the known case study:
the relational database theory for plain relations,
frame theory for potential objects.

The implementation results in the non-plain indexed relations
and gives the generalized notion of the (relational)
database management system.
To extract its possible advantages the
unified architecture based on extensible data objects' model is proposed.
It supports the main object-oriented mechanisms of encapsulation,
polymorphism and inheritance and contains five major components.
They are
Conceptual Shell (CS), Application Development Environment (ADE), Basic
Relational Tool System (BRTS), Meta Relational Tool System (MRTS),
and System Excnage.
Some general counterparts neatly corresponding to ADE are to be
reflected as a two layered scheme.
The first layer consists of the relational sybsystem
and the second layer consists of the frame-based event
driven subsystem. All the operations of the higher level
are mapped into the corresponding operations within
the lower level.

Both the subsystems are relatively self-contained, and
incorporated into the ToolKit. BRTS supports relational interfaces
({\em extentional level}), MRTS adds the {\em intentional level},
and the pair $<$BRTS,~MRTS$>$ gives the enhanced relational
features. ADE contains the extensible data objects' model and
manipulates the switching, or {\em variable concepts} so that the
computational idead of functor-as-object is resolved.

Both BRTS and MRTS are
embedded into ADE. CS adds an external interactive mode to maintain and
generate applications.

\section{The relative approaches in use}\label{Section:3}

The ADE notion is based on the theory of variable concepts in the form of
{\em variable sets}.
The origin of the variable sets would be commented
taking in mind the computations in Figure~\ref{pic:4}.
The resulting evaluation according to the commutative diagram
generates the types (intensional objects) which
are determined by the indexed sets (extentional objects).
This assignment, as we see, depends on a category
of indices in use.

The same constructions were used in an elementary category theory,
in particular in representation of predicate logic.
Until present time these results were applied mainly
to the semantics of programs especially in denotational semantics of
programming languages.
Here the construction in Figure~\ref{pic:4}
fills in the gap between evaluating the expressions
in a category and generating the derived types in a database
environment.

The attempts to formalize an idea
of `information system'
\cite{Article:93:Rishe:DataTrans}
with the lattices of the domains
as a basis of smart computations are known.
However, their fruitfulness is observed
for very special and restricted cases.

The logical attempts to investigate the
concepts of the `object' and the `substitution' in order
to render theoretical
computer science with the trustworthy representations
are known
\cite{Article:94:Thalheim:DBDesign}.
The indicated difficulties have the same nature as
in a predicate calculus.

The results obtained by \cite{Article:95:Burke:QueryByNavigation},
give the background to combine databases and knowledge bases for assisted
browsing. In particular, the contribution of the current
research is to represent `stages of knowledge' by the transitions
observed in the commutative diagrams.

All these investigations as well as adjoining contemporary works
employ the idea of `object' borrowed from more or less traditional
ideas of the object-oriented approach.

The analysis of problems arose and, in particular, representation of
transition processes in information systems  reflects
the principle effect of varying the primary established objects.
The attempts
of partial solutions do not yet result in an adequate
mathematical apparatus.

\section{The relative applied research}\label{Section:4}

The starting point for ADE project originates from the
various object notions.

Known results in the field of relational data bases,
in the field of conceptual
modelling, in the field of knowledge based
systems (including inference and transformation mechanisms) chiefly deal
with established modes of information systems.
For this situation an assumption of implicite environment
which is fixed before the analysis is prescribed.
Nevertheless an update of the database usually violate
the consistency of the data model.

Allowing different transaction mechanisms brings more
flexibility to the data model. But this potential flexibility
to be implemented needs some sophisticated and difficult
in implementation mechanisms with very complicated
computational background.

An information system developer needs the transparent
solutions based on some clear to understood conceptions.
The main advatage of disadvantage of the practical solution
depends on the initial `building blocks'. The known solutions
tends to maintain some kind of framework which ramify
the behaviour of the objects.

Establishing the representation of a concept as invariant on possible
changes, creating the tools of `evolving the concept'
as a mathematical process
overtakes these essential difficulties.
The distinction between potential and actual objects
makes some contribution to
an information system dynamics, and, in particular,
to better understanding of the environment
transition, or switching processes
whenever the knowledge concerning the objects
and their behaviour drifts from one `stage of knowledge'
to another `stage of knowledge'.

Developing the derived types' evaluator as in Figure~\ref{pic:4}
is also important by the following reasons.
Information systems contain besides database (DB) also the metadata base
(MDB) that does include different facts concerning database.
These fact are mainly related, as we see, to the
domains of potential objects. The potential objects
are to be observed as the `pieces of knowledge'
and are naturally the {\em metadata}.

In applications metadata bases are
often used as the components of knowledge system with
knowledge bases. Both parts, DB and MDB communicate each other and the
host computational environment of information system.
Some incite concerning their shared functions would
be given by the commutatiive diagram to ealuate
e.g., the derived types.

The relative ideas were understood from 1987 in the nine known research
projects of {\em extensible} database systems.
The main
ideas were concentrated on a model of data. Even the project without any
pre-defined data model was proclaimed, but the desirable degree of
flexibility was not achieved. The most prominent experiments with
experimental (object-oriented) extensible system have used the molecular
chaining of objects to avoid the known difficulties in designing and
development. But the completed solution was not obtained neither in a theory
nor in practice.

Some kind of contribution to override this problem is given here
by the Modular Counterparts (MC).

The design of the systems which accumulate and process both
information and metainformation does influence the effectiveness of
information system, the efforts of its designing, maintenance, modifications
and extentions. This results in the problem of designing the Application
Development Environment (ADE) which communicates with the varying DB
and MDB.

\section{Objectives of the proposed ADE conception}\label{Section:5}

ADE is both the research project and tool kit aimed at development of
object evaluator which is based on the sound
computational ground.
The computational models and methods which being considered
together comprise a theory of computations for object-oriented
information systems which covers both the data management
and knowledge based solutions.
In particular, the knowledge concerning the possible
behaviuor of the {\em actual} domains is contained in
the corresponding {\em potential} domains. The measure
of correspondence is mainly determined by the details
of {\em indexing} procedure.

The main research aims denominated here are to
bring more meaning in evaluations with the objects.
Some dominant trends are to persue the knowledge
of the entities as follows:
data object to represent the
storable data; metadata object to represent the conceptual information;
assignment to represent internal states and dynamics; extensible database to
represent relatively self-contained couples of objects and some additional
derived concepts.

All of these notions are important to succeed in ADE development.
At the present time
in a field of information systems' development
the experience having been accumulated deals with
the methods to handle static or quasi-static application domains.

The mathematical approvements, e.g., database {\em consistency}
are adopted almost exclusively for static domains.

Swithching the computational environment
brings in both the flexibility and unexpected difficulties.
For instance, the data object understanding is now based on
the higher order logic. The corresponding data model
is assosiated with the higher order function spaces.
We fill that the information system community
has the minimal experience to establish and analize
the corresponding data models.

The feasibility of the higher order object evaluator
is another point of interest that influenced the
current research. Our proposal is to use the commutative
diagram techique, at least, whenever other models fail
to support the object evaluation.

\section{ADE contribution to the research activity}\label{Section:6}

ADE has under research the idea of a concept as the variable entity to
possess the creation of the variable concepts and associated transition
effects.
In their turn the variable concepts lead to parameterized type system. The
approach developed in ADE is based on the reasons stated.
     The usage of the method of embedding typed system (including the
apparatus of variable concepts) into untyped system
is based on the commutative diagrams support.
     Combining the ideas of variable concepts will make possible
development of a wide range of applied information systems, particularly in
the field of data base management systems
for advanced applications whenever the knowledge
fragments are explicitely or implicitely involved.

\section{General features}\label{Section:7}

\begin{figure*}
\unitlength 1.00mm \linethickness{0.4pt}
\begin{picture}(133.00,122.00)
\multiput(80.00,55.00)(0.89,-0.09){3}{\line(1,0){0.89}}
\multiput(82.66,54.73)(0.36,-0.12){7}{\line(1,0){0.36}}
\multiput(85.21,53.91)(0.21,-0.12){11}{\line(1,0){0.21}}
\multiput(87.53,52.60)(0.13,-0.12){15}{\line(1,0){0.13}}
\multiput(89.54,50.83)(0.11,-0.15){14}{\line(0,-1){0.15}}
\multiput(91.15,48.69)(0.11,-0.24){10}{\line(0,-1){0.24}}
\multiput(92.28,46.27)(0.10,-0.43){6}{\line(0,-1){0.43}}
\put(92.89,43.67){\line(0,-1){2.67}}
\multiput(92.96,41.00)(-0.12,-0.66){4}{\line(0,-1){0.66}}
\multiput(92.48,38.37)(-0.11,-0.28){9}{\line(0,-1){0.28}}
\multiput(91.47,35.89)(-0.11,-0.17){13}{\line(0,-1){0.17}}
\multiput(89.98,33.67)(-0.12,-0.12){16}{\line(-1,0){0.12}}
\multiput(88.07,31.81)(-0.19,-0.12){12}{\line(-1,0){0.19}}
\multiput(85.81,30.37)(-0.31,-0.12){8}{\line(-1,0){0.31}}
\multiput(83.31,29.43)(-0.66,-0.10){4}{\line(-1,0){0.66}}
\multiput(80.67,29.02)(-1.33,0.07){2}{\line(-1,0){1.33}}
\multiput(78.00,29.15)(-0.43,0.11){6}{\line(-1,0){0.43}}
\multiput(75.41,29.84)(-0.24,0.12){10}{\line(-1,0){0.24}}
\multiput(73.02,31.03)(-0.15,0.12){14}{\line(-1,0){0.15}}
\multiput(70.93,32.69)(-0.11,0.14){15}{\line(0,1){0.14}}
\multiput(69.21,34.74)(-0.11,0.21){11}{\line(0,1){0.21}}
\multiput(67.96,37.10)(-0.11,0.37){7}{\line(0,1){0.37}}
\multiput(67.21,39.67)(-0.10,1.33){2}{\line(0,1){1.33}}
\multiput(67.00,42.33)(0.11,0.88){3}{\line(0,1){0.88}}
\multiput(67.35,44.99)(0.11,0.32){8}{\line(0,1){0.32}}
\multiput(68.23,47.51)(0.11,0.19){12}{\line(0,1){0.19}}
\multiput(69.60,49.80)(0.11,0.12){16}{\line(0,1){0.12}}
\multiput(71.42,51.76)(0.17,0.12){13}{\line(1,0){0.17}}
\multiput(73.60,53.31)(0.27,0.12){9}{\line(1,0){0.27}}
\multiput(76.05,54.38)(0.66,0.10){6}{\line(1,0){0.66}}
\put(80.00,46.00){\makebox(0,0)[cc]{{\bf S}ystem}}
\put(80.00,38.00){\makebox(0,0)[cc]{{\bf E}xchange}}
\put(68.00,46.00){\vector(-1,0){8.00}}
\put(60.00,38.00){\vector(1,0){8.00}}
\put(92.00,38.00){\vector(1,0){8.00}}
\put(100.00,46.00){\vector(-1,0){8.00}}
\put(30.00,22.00){\dashbox{3.00}(30.00,40.00)[cc]{ }}
\put(45.00,52.00){\oval(26.00,10.00)[]}
\put(45.00,32.00){\oval(26.00,10.00)[]}
\put(40.00,47.00){\vector(0,-1){10.00}}
\put(50.00,37.00){\vector(0,1){10.00}}
\put(100.00,22.00){\dashbox{3.00}(30.00,40.00)[cc]{ }}
\put(115.00,52.00){\oval(26.00,10.00)[]}
\put(115.00,32.00){\oval(26.00,10.00)[]}
\put(110.00,47.00){\vector(0,-1){10.00}}
\put(120.00,37.00){\vector(0,1){10.00}}
\put(45.00,52.00){\makebox(0,0)[cc]{{\bf PO}-library}}
\put(45.00,32.00){\makebox(0,0)[cc]{{\bf PO}-relations}}
\put(115.00,52.00){\makebox(0,0)[cc]{{\bf AO}-library}}
\put(115.00,32.00){\makebox(0,0)[cc]{{\bf AO}-relations}}
\put(76.00,30.00){\vector(0,-1){13.00}}
\put(84.00,17.00){\vector(0,1){13.00}}
\put(63.00,9.00){\oval(26.00,10.00)[]}
\put(97.00,9.00){\oval(26.00,10.00)[]}
\put(48.00,1.00){\dashbox{3.00}(64.00,16.00)[cc]{ }}
\put(62.00,11.00){\makebox(0,0)[cc]{Retrieval}}
\put(62.00,6.00){\makebox(0,0)[cc]{System}}
\put(97.00,11.00){\makebox(0,0)[cc]{Storage}}
\put(97.00,6.00){\makebox(0,0)[cc]{System}}
\put(75.00,12.00){\vector(1,0){10.00}}
\put(85.00,6.00){\vector(-1,0){10.00}}
\put(76.00,54.00){\vector(0,1){13.00}}
\put(84.00,67.00){\vector(0,-1){13.00}}
\put(115.00,79.00){\oval(26.00,10.00)[]}
\put(45.00,79.00){\oval(26.00,10.00)[]}
\put(80.00,79.00){\oval(26.00,10.00)[]}
\put(44.00,79.00){\makebox(0,0)[cc]{{\bf PO}-manager}}
\put(80.00,79.00){\makebox(0,0)[cc]{{\bf DPO}-manager}}
\put(115.00,81.00){\makebox(0,0)[cc]{{\bf R}elational}}
\put(115.00,76.00){\makebox(0,0)[cc]{manager}}
\put(40.00,67.00){\vector(0,1){7.00}}
\put(50.00,74.00){\vector(0,-1){7.00}}
\put(72.00,67.00){\vector(0,1){7.00}}
\put(87.00,74.00){\vector(0,-1){7.00}}
\put(110.00,67.00){\vector(0,1){7.00}}
\put(120.00,74.00){\vector(0,-1){7.00}}
\put(40.00,67.00){\line(1,0){80.00}}
\put(40.00,91.00){\vector(0,-1){7.00}}
\put(50.00,84.00){\vector(0,1){7.00}}
\put(72.00,91.00){\vector(0,-1){7.00}}
\put(87.00,84.00){\vector(0,1){7.00}}
\put(110.00,91.00){\vector(0,-1){7.00}}
\put(120.00,84.00){\vector(0,1){7.00}}
\put(40.00,91.00){\line(1,0){80.00}}
\put(36.00,97.00){\framebox(89.00,10.00)[cc]{{\bf CS}:\ \ \ {\bf
C}onceptual {\bf S}hell }} \put(76.00,97.00){\vector(0,-1){6.00}}
\put(84.00,91.00){\vector(0,1){6.00}}
\put(115.00,117.00){\oval(26.00,10.00)[]}
\put(45.00,117.00){\oval(26.00,10.00)[]}
\put(80.00,117.00){\oval(26.00,10.00)[]}
\put(43.00,112.00){\vector(0,-1){5.00}}
\put(47.00,107.00){\vector(0,1){5.00}}
\put(78.00,112.00){\vector(0,-1){5.00}}
\put(82.00,107.00){\vector(0,1){5.00}}
\put(113.00,112.00){\vector(0,-1){5.00}}
\put(117.00,107.00){\vector(0,1){5.00}}
\put(45.00,117.00){\makebox(0,0)[cc]{{\bf C}-interface}}
\put(80.00,117.00){\makebox(0,0)[cc]{{\bf DPO}-interface}}
\put(15.00,117.00){\makebox(0,0)[cc]{Interface:}}
\put(15.00,102.00){\makebox(0,0)[cc]{Shell:}}
\put(15.00,79.00){\makebox(0,0)[cc]{Manager:}}
\put(15.00,42.00){\makebox(0,0)[cc]{Object:}}
\put(15.00,9.00){\makebox(0,0)[cc]{System:}}
\put(115.00,117.00){\makebox(0,0)[cc]{{\bf CR}-interface}}
\put(27.00,19.00){\dashbox{2.00}(36.00,75.00)[cc]{ }}
\put(97.00,19.00){\dashbox{2.00}(36.00,75.00)[cc]{ }}
\put(127.00,91.00){\makebox(0,0)[cc]{\bf BRTS}}
\put(33.00,91.00){\makebox(0,0)[cc]{\bf MRTS}}
\put(43.00,97.00){\vector(0,-1){3.00}}
\put(47.00,94.00){\vector(0,1){3.00}}
\put(113.00,97.00){\vector(0,-1){3.00}}
\put(117.00,94.00){\vector(0,1){3.00}}
\end{picture}
\caption{\sf {\bf A}pplication {\bf D}evelopment {\bf
E}nvironment: {\bf ADE}. {\small     (Abbreviations:
         {\bf C}  - {\bf C}onceptual;
         {\bf CR}  - {\bf C}onceptual {\bf R}elational;
         {\bf DPO} - {\bf D}escendent {\bf P}otential {\bf O}bject;
         {\bf CS}  - {\bf C}onceptual {\bf S}hell;
         {\bf PO}  - {\bf P}otential {\bf O}bject;
         {\bf AO}  - {\bf A}ctual {\bf O}bject;
         {\bf MRTS} - {\bf M}eta {\bf R}elational {\bf T}ool {\bf S}ystem;
         {\bf BRTS} - {\bf B}asic {\bf R}elational {\bf T}ool {\bf S}ystem.
{\bf C}onceptual interface gives the abilities to establish, update
or remove the restrictions imposed to the objects in use.
The objects are `based'. The {\bf PO}-base consists of the
potential objects and is equipped with the language to maintain
their features. They are conformed with a {\em partial order}
which enables the {\em inheritance} of the properties. The atomic
{\bf PO} are setted up as the {\em sorts}. {\bf DPO}-interface
serves the set-theoretic operations, triggers and built-in
procedures on sorts.
{\bf AO}-library contains the {\em derived}
types which are used as ranges for the relational {\bf S}earch {\bf S}ystem.
To maintain the {\em indexing} procedure {\bf DPO}-interface
is script-based and use the SQL-cursors.
{\bf CR}-interface is attached to the {\bf AO}-base.
{\bf S}ystem {\bf E}xcange makes a valid request and response for the
system counterparts.)
}
}\label{pic:2}
\end{figure*}
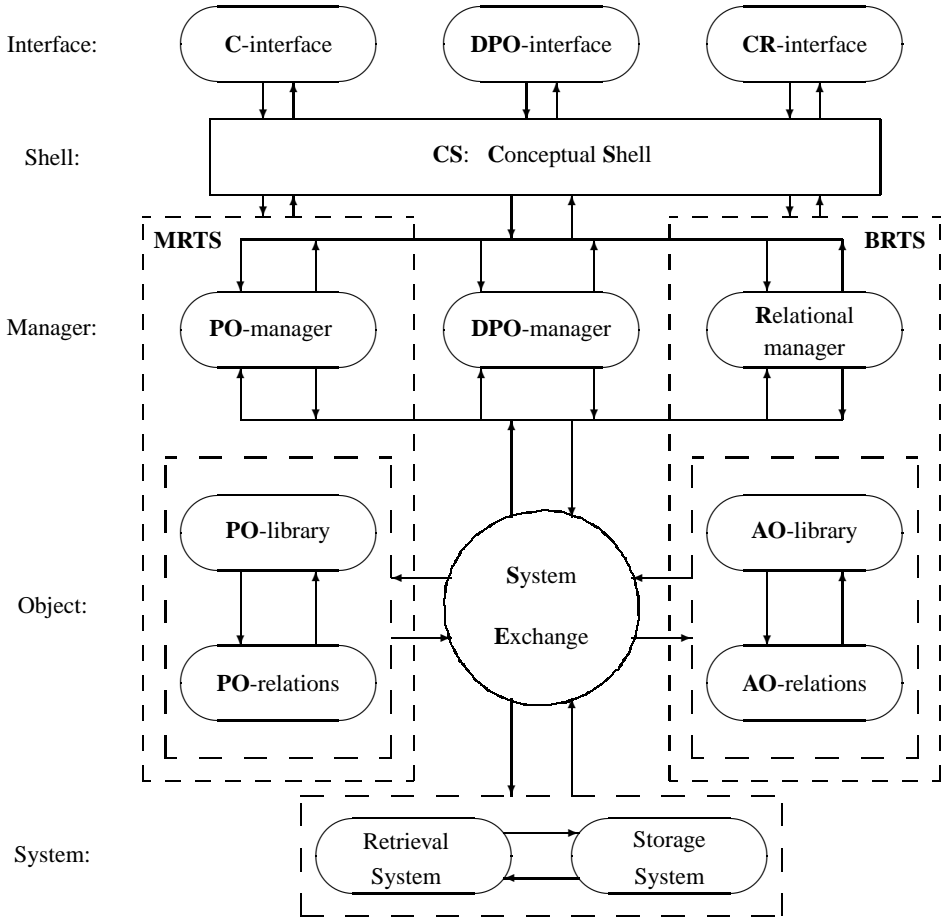

ADE is implemented on the base of the two layered architecture
(shown in Figure~\ref{pic:2}) and
is viewed to be a comprehensive research
as follows:

establishing the primitive frame to represent and analyze a `variable
concept';

setting up the approach to integrate the
commutative diagram technique with the needs
of evaluating the objects;

developing methods to adopt some intentional concepts,
in particular, to develop and use corresponce between
potential and actual objects;

creating the tool kit to explicate and apply the advantages of variable
concepts and generating the derived types `on fly';

augmenting the possibilities of host programming system
to fit in the demands of switching the computational environment;

specifying the enhanced data models;

fixing the possible ranges of design and development those systems that
involve the idea of data/metadata object;

creating the generalized tool kit on the basis of the mathematical concepts
in the current contribution.

The target prototype system Application Development Environment (ADE) is
mainly based on the idea of variable, or switching concept and covers the
vital mechanisms of encapsulation, inheritance and polymorphism. Variable
concepts naturally generate families of similar types that are derived
from the
generic types, or sorts.
Concepts in ADE are equipped with the evolvents that manage
the transitions, or switching between the types. In particular, the identity
evolvent supports the constant concepts and types (statical concepts).
To achieve the needed flexibilty a general ADE layout consists of
the uniform modular units, as shown
in Figure~\ref{pic:2}.

In ADE Data Object Definition Language (DODL) contains the construction
of data objects' base scheme as a relation between concepts. Concepts are
included into the type system with the interpretation over the variable
domains. A coherent set of variable domains generates the data objects'
base.
Basis to maintain the data objects in use and their bases is generated by
computational models with applicative structures.
The developer obtains the set of the means that establish, support and modify
the linkages between the data objects' base schemes, data objects' base and
computational models. DODL declares: {\em type system} as a set of metadata
objects; em linkages between the types; system of domains;
linkages between the
domains; extentions of domains and types; computational tools of applicative
pre-structures and structures.

The third part of the implementation supports two level of interfaces. The
first is the Intentional Management System (IMS) to support concepts
(metadata objects) of different kinds, and the second is associated Extentional
Management System (EMS) to support the appropriated extentions (data
objects) generated by the intentions. Both the counterpats are included
into {\bf C}onceptual {\bf Shell}.

EMS is embedded into the unified computational model. It is object-oriented
extensible programming system
{\bf B}asic {\bf R}elational {\bf T}ool {\bf S}ystem (BRTS). BRTS
has the fixed architecture with the one level comprehension, separate self-
contained components, interfaces and languages. It is the First Order Tool
(FOT) and generates `fast prototypes'. D(M)ODL and D(M)OML of BRTS
contain the SQL-based relational complete languages
that cooperate with ADE.
BRTS mainly supports relatively large number of low cardinality relations
(extentions) and supports Data (Metadata) Object Model D(M)OM with
retrieval, modifications and definitions of a storable information.

IMS is also embedded into the computational model and supports a numerous
matadata objects. Their amount is almost the same as for data objects. IMS
is based on D(M)OM with a simple comprehension to manage metadata base
and is supported by
{\bf M}eta {\bf R}elational {\bf T}ool {\bf S}ystem (MRTS). MRTS
manipulates with the metaobjects (concepts) and metarelations (frames) and
is embedded into ADE.

\section{New supporting technologies}\label{Section:8}

A main result is the experimental verification of variable concepts approach.
This would be applied to develop the variety of applied information systems.

Computations with variable concepts and appropriated programming system
allows to built a system especially to manipulate the objects.
One of the features enables declarations and manipulations
of superobjects.

The database
extensibility, development of databases with varying data models,
dynamical
databases, analysers of `data object' dynamics,
the `bases of invariants' etc.
would be achieved.

The knowledge base systems grounded on the concepts which are metadata
objects. This feature is based on the {\bf PO} techniques.
Maintenance of the systems of `variable concepts' and their
interrelations, management of switching the systems of concepts,
management of database modification etc.

\section{Usage of intentions and extentions}\label{Section:9}

ADE gives a smart framework for intentions and extensions and intentional
tool/applied system. It enables the possibility to develop the conceptual
support that encircles the ajunct ideas:

establishing the logical apparatus
(on the basis of higher order theory) to
study the hierarchies of variable concepts;

development of specialized `tool theories' to estimate a selective power of
newly designed programming systems, database systems and models for
systems with databases and knowledge bases;

support and development of specialized semantical theories and models for
systems with databases and knowledge bases.

Possible applications of this framework: `rapid prototypes' of newly
developed computer information systems, estimation of their ranges,
adaptation to variations in problem domains, demands, programming
systems, experts. The specific feature of architecture: two levelled design -
intentional and extentional levels.

\section{Improvements in design, engineering and management processes}
           \label{Section:10}

At present it is difficult to estimate real benefits of using
the apparatus of
variable concepts in particular applied information systems.
This is due to current study of the
mathematical apparatus being developed and
the approach on the whole. The possible gains and prospects that may be
yielded by the ADE approach are set forth below:

high degree of generalization of the approach developed and therefore the
unification of the tool system design;

taking into account the intentional features (`subjects') within the
mathematics in development and creation of `dynamical scripts' for the
problem domain;

relative simplicity of the resulting formalization and adequate languages of
interfaces;

possibility to aggregate/disaggregate the representations of the entities and
feasible and transparent tolls to capture more meaning of the objects;

clarity, referential transparency and fully explicit constructions that are in
the nature of relational systems and simplification both of
system and applied programming by the compositional, or
functional style;

possibilities to handle collections of objects or concepts;

establishing of formal object properties inheritance systems and its
modular implementation;

potential possibilities of straightforward exploiting the apparatus of higher
order logic (including the descriptions theory)  in an
area of information systems to generate the enhanced means and tools of
design, implementation and management with a powerful selectivity.

\section{Technical description}\label{Section:11}

\subsection{Experimental techniques}

The feature of current research is in primary
creation of the needed tool system to be adequate to the newly generated
mathematical apparatus.

{\bf Conceptual Shell}.
An approach to provide experiments needs the
technological object-oriented {\bf C}onceptual {\bf S}hell (CS).
CS implementation
gives a systematical approach to development, application and use of the
programming systems with databases and metadata bases to support:

variations occurred in a problem domain or in its representation,
especially the more detailed or aggregated information;

database and metadata base management system to integrate data objects,
metadata objects and programs within the unified computational framework -
Extensible Computational Environment;

designing the representation of problem domain by linking data objects
and metadata objects;

growth of a variety of datum, metadatum and corresponding languages.

The main counterparts of the project are briefly summarized below.

{\bf Application Development Environment}.
The basic computational
mechanisms that support evaluation of the expressions, switching the
concepts, declaring of the descriptions for intentions, generating their
extentions and extensibility of primary data object model.

{\bf Basic Relational Tool System}.
Supports the purely relational and/or
enhanced solutions to validate the storable data objects (extentional level).

{\bf Meta Relational Tool System}.
Supports the relational solutions for
metadata objects with the inheritance of properties (intentional level).
The important task of project would be resolved when both BRTS and
MRTS will be embedded into ADE. This enables the possibility of higher
order logic to the direct experimental verification.

{\bf System Exchange}. Supports the correspondence between
the potential and actual obkects in use. Generates the
linkages between the {\bf APO}-objects and {\bf DPO}-objects.

\subsection{Software}

The experimental research and verification of the obtained
model is based on prototypes - {\bf CS, ADE, BRTS, BMRS}.
The difficulties to
implement full scale prototype are resolved by the high level object-oriented
programming language. Some candidate programming systems are tested to
enable the needed computational properties. After that the main
programming tool kit is selected. Preliminary candidate tools were C++ or
Modula-2. An attention is paid to select an appropriate database management
system. If needed the original DBMS is attached. At the preliminary tests the
attention was paid to OLE-2 techniques.

\subsection{Tools}

Some ready made original systems were tested and expanded to
achieve the prototype system with the properties mentioned.

\subsection{Generation of the results and applications}

This is multistage process as follows:
1) creating the data objects' base; 2) creating the metadata objects' base;
3) embedding the results of the previous stages into ADE with subsequent
extentions and adding the dynamics; 4) generation of concepts' families;
5) switching the families; 6) performance analysis and improvements. An
important faze is gathering the data and generating the data objects,
development of applied data objects'/metaobjects' bases. This enables the
extentional/intentional level of information in use to apply the enhanced
methods of higher order (relational) logic.

\subsection{Evaluation of the results}

The basic step of evaluation is to generate
the known results in the information systems technology. The most
prominent particular cases are listed below.

{\bf Relational solutions}.
Are achieved by BRTS (the first order
extentions). These are the tunes of relational completeness within the first
order relational language (the basis of induction). The step of induction by
complexity (the higher order extentions) is in iterating the first order
relational structures to obtain the higher order structure. This is achieved by
embedding BRTS into ADE with the subsequential experiments to verify
possible restrictions of implementation.

{\bf Meta relational solutions}.
They are imposed by the intentions of the objects and correspondences
between potential and actual objects.
Are directly supported by MRTS (the
first order intentions) including the inheritance of properties. This supports
the basis of induction. The step of induction generated the composite
intentions (the higher order intentions). The observation is that the concepts
in this sense are exactly the statical concepts from the proposed theory of
variable concepts. Therefore, embedding MRTS into ADE with the statical
restrictions of the concepts generate the first order frame (or intentional)
system without comprehenced frames.

{\bf Awaited corollary}.
Restriction of the variable concepts to the statical
(identity) evolvents enables the known (meta)realtional solutions. The two
level architecture is adopted with the  $<$intention,~extention$>$ pair
matching the traditional representation of (meta)data objects.

{\bf Enhanced properties}.
The potential (or actual) objects can be composed to result in
the valid expressions. These expressions are evaluated via the
{\bf C}onceptual {\bf S}hell and the series of transmissions
via {\bf S}ystem {\bf E}xchange. The events are viewed
as the interrelated entities. A natural assumption is the
possibility to compose distinct events generating the
{\em evolvents of events}.

{\em Non identity evolvents}
 generate the families of
ordinary concepts and capture the dynamics of evaluations. The verification
of this effect (`switching concepts') is included in the final stage of the
project and needs a careful selection of the problem domains that clearly and
transparently demonstrate all the dynamical features and advantages of the
method. Additional experiments will include the `partial' and `fully'
dynamical problem domains with the appropriated applied information
systems.

{\bf Awaited corollary}.
The dynamical properties are captured by both
the intentions (`weak dynamics') and the evolvents (`strong dynamics').
Upon finishing the (full scale) prototype system, its implementation and
verification the higher order object oriented information system (tool and
applied) with the dynamical properties (and switching concepts) will be
achieved.

\section*{Conclusions: interpretation of the results}
\addcontentsline{toc}{section}{Conclusions}

The resulting two level comprehension model and computational environment
verify the feasibility of the approach. The adequate, neutral and semantical
representation of data is the target in the sphere of extensible systems and
their moderations and modifications.
The relational solutions are the criteria in database technology.
Therefore, the
variable concepts generate the power and sound representation of data
objects, have the boundary conditions as the known results in information
systems (both in a theory and applications) and capture the additional effects
of dynamics to simulate, in particular, the encapsulation, polymorphism and
inheritance. The last gives the contribution in development of object-oriented
systems.


\addcontentsline{toc}{section}{References}
\newcommand{\etalchar}[1]{$^{#1}$}


\end{document}